# Analytical Modeling of $k_{33}$ Mode Partial Electrode Configuration for Loss Characterization

Yoonsang Park[1,*], Maryam Majzoubi[1], Yuxuan Zhang[1], Timo Scholehwar[2], Eberhard Hennig[2], and Kenji Uchino[1]

1) International Center for Actuators and Transducers (ICAT), The Pennsylvania State University, University Park, PA, 16802, USA

2) R&D Department, PI Ceramics GmbH, Lindenstrasse, 07589 Lederhose, Germany

**Abstract:** Accurate determination of three types of losses (dielectric, elastic and piezoelectric) in piezoelectric materials is critical, since they are closely related to the performance of high-power piezoelectric devices. The Standard $k_{33}$ mode has a number of serious deficits that hinder researchers from determining accurate physical parameters and losses. In order to overcome such deficits, "partial electrode" has been devised and proposed. This study provides analytical derivation process and proposes parameter determination method by utilizing analytical solutions. Compared to finite element analysis, analytical solutions show 0.1 % difference in resonance frequencies and 2 % difference in mechanical quality factors, proving themselves as valid modeling. The analytical solutions are fitted to experimental data to determine physical parameters and losses. The $k_{33}$ (electromechanical coupling factor) values were calculated with determined values from curve fitting in two different ways and show good agreement to each other.

## I. Introduction

For the past decades, piezoelectric materials have been proven to be better than electromagnetic (EM) counterparts when being utilized as ultrasonic motors, sonar transducer devices and electrical transformers, since they have higher energy densities in smaller volumes and generates no electromagnetic noise. Accurate loss determination of piezoelectric material is crucial for the following two reasons: First, the bottleneck of high-power piezoelectric devices has been identified as heat generation, which severely degrades their performances [1-2]. This heat generation can be explained by three types of losses: elastic ($\tan\phi$), dielectric ($\tan\delta$), and piezoelectric losses ($\tan\theta$). Second, finite element analysis has become imperative for devising and optimizing piezoelectric devices [3-5]. Exact values of intensive/extensive parameters and losses are required for the accurate simulation results of finite element analysis, especially near resonance frequencies. Therefore, it is important to understand three types of losses in different vibration modes, as well as elucidate loss mechanism.

Each dielectric, elastic, and piezoelectric loss is further classified into "intensive" and "extensive" subsets according to the constraint conditions of the vibration modes: intensive parameters, such as stress ($X$) and electric field ($E$), are the ones that rely upon the amount of matter or total volume of the system, whereas extensive parameters, such as strain ($x$) and dielectric displacement ($D$), are the ones that depend on the volume of given system. Followed by the statement of intensive and extensive parameters, the following equations are the definitions of intensive and extensive losses [6]: $\varepsilon^{X*} = \varepsilon^{X}(1 - j\tan\delta')$, $s^{E*} = s^{E}(1 - j\tan\phi')$, $d^{*} = d(1 - j\tan\theta')$; $\kappa^{x*} = \kappa^{x}(1 + j\tan\delta)$, $c^{D*} = c^{D}(1 + j\tan\phi)$, $h^{*} = h(1 + j\tan\theta)$.

The losses can be represented as tangent angle supposing that the loss values are not large and handled primarily as viscous friction from the viewpoint of mathematical simplicity.

The first three equations explain intensive losses, which are primed; $\varepsilon^X$, $s^E$ and $d$ are relative permittivity under constant $X$, elastic compliance under constant $E$ and piezoelectric constant, respectively. Primed losses are defined as losses under constant intensive parameters. The latter three equations explain extensive losses, which are non-primed; $\kappa^x$, $c^D$ and $h$ are inverse permittivity under constant $x$, elastic stiffness under constant $D$ and inverse piezoelectric constant, respectively. Therefore, the non-primed losses are defined as losses under constant extensive parameters.

Up to date, Institute of Electrical and Electronics Engineering (IEEE) have proposed a standard method for parameter determination with different vibration modes [7]. One of the problems with IEEE Standard is that it introduces only one type of elastic loss, which results in mechanical quality factor at resonance frequency ($Q_A$) is equal to that at antiresonance ($Q_B$). However, there exist significant divergence with the experimental data (e.g. $Q_A < Q_B$ in PZT [8]). This issue was resolved by explicitly introducing the "piezoelectric loss". Another issue of IEEE is a limited parameter determination [7,9]. For example, the IEEE Standard $k_{31}$ mode specimen can only provide intensive parameters [9], while the Standard $k_{33}$ or $k_t$ specimen can exclusively provide "extensive" elastic parameters [10], causing error accumulation for obtaining intensive elastic parameters from the transformation formulas. In addition, the Standard $k_{33}$ mode also suffers from small capacitance and fringing electric field that are originated from its intrinsic structure [11-13].

In order to resolve such issues, partial electrode (PE) configuration method was introduced to realize the mechanical excitation measurement, in contrast to the "Standard" specimen with pure electrical excitation. PE configuration sample has two main parts: center part, which is electrically excited and behaves as a mechanical "actuator", and side part, which is pure mechanical load excited by the center part and can be modified with electrical constraints. In our previous paper [9], the $k_{31}$ mode PE was designed in order to experimentally determine both intensive and extensive elastic compliances and losses. It enables direct determination of extensive elastic parameters with non-electrode (NE) specimen, significantly reducing the error for extensive elastic parameters. For the $k_{33}$ mode, the PE design, as shown in Figure 1, are composed of the open circuit (OC) side load, which corresponds to antiresonance drive of the sample, short circuit (SC) load, which corresponds to resonance drive, and side electrode (SE), which enables direct measurement of "intensive" elastic compliance and loss. The PE configuration for $k_{33}$ mode gives great degree of benefits: it resolves small capacitance issue in the standard $k_{33}$ mode specimen because the center part gives 100 – 200 higher capacitance values than the standard specimen. Also, the users can avoid specimen setup issue. When the sample is clamped in the vibration direction on the Standard $k_{33}$ specimen, a distortion of admittance spectrum can happen due to the sample vibration constraint [11,14]. Furthermore, due to the length consideration of the side part, the parameter distortion effect due to fringing electric field is smaller than the standard sample [11], and PE enable direct determination of intensive elastic properties by putting electrode on the side.

In this paper, we provide detailed derivation process for analytical solutions for $k_{33}$ mode PE configuration for providing the determination process of the intensive and extensive piezoelectric parameters and corresponding losses. The derived admittance equations are validated by the comparison with finite element analysis (FEA) and experimental admittance curves.

## II. Methods

In order to solve analytical solutions and generate admittance curves for $k_{33}$ mode PE configuration, Wolfram Mathematica (version 11.1, Wolfram Research, Champaign, IL) was utilized. The FEA software package (ATILA++ distributed by Micromechatronics, Inc.,

State College, PA) was used to validate admittance curves from analytical solutions, which can integrate all kinds of losses, including piezoelectric loss, in the simulation. The comparison was done between admittance curves from ATILA FEA and analytical solutions with the exactly same input parameters. Also, all samples, PE OC, PE SC and PE SE, were prepared and measured with Precision Impedance Analyzer (4294A, Agilent Technology, Santa Clara, CA) with constant peak voltage of 0.1 mV to avoid possible peak distortion [15]. In order to determine the $k_{33}$ mode related parameters for the PE admittance curve fitting, the standard $k_{31}$ sample were prepared in parallel and measured with the same impedance analyzer for precisely determining the center part's physical parameters in advance. The physical parameters and losses were determined as average, and errors were determined with error propagation method. For piezoelectric material, PIC 255 (PI Ceramics GmbH, Lederhose, Germany), soft PZT with morphotropic phase boundary composition, was used and Au was sputtered to create electrodes. The bulk ceramic block was poled and cut to create plate with the dimension of 20 × 2.5 × 0.5 mm (poled along 20 mm direction). Then, the center of the plate was electrode and re-poled along 0.5 mm direction. The measured admittance curves then fitted with analytical solutions in order to determine related physical parameters.

III. Derivation of Analytical Solutions for Partial Electrode Configuration

A. General and Mechanical Wave Equations

Consider the following geometry for PE configuration, as shown in Figure 2. The effective coordinate is defined as shown. The $x$, $y$ and $z$ direction are along with the configuration's length ($l$), width ($w$), and thickness ($t$) direction, respectively. The numbers in the parenthesis in the axis correspond with the indices of physical parameters that are subscripted. The portion of center electrode is denoted as "*a*" (0 < *a* < 1). The zero coordinate is set at the center of the bar for convenience, since the configuration is exactly symmetrical in $x$ direction. It should be noted that this analysis is based on 1D assumption that $l \gg w, t$ and only vibration along the length is considered. Note that the slight discrepancy with the 3D ATILA simulation results will be attributed from this simplified assumption. The piezoelectric constitutive equations for $k_{33}$ mode PE are given by:

$$x_{1c} = s_{11}^E X_{1c} + d_{31} E_{3c} \tag{7}$$

$$D_{3c} = d_{31} X_{1c} + \varepsilon_0 \varepsilon_{33}^X E_{3c} \tag{8}$$

$$x_{1s} = s_{33}^E X_{1s} + d_{33} E_{1s} \tag{9}$$

$$D_{1s} = d_{33} X_{1s} + \varepsilon_0 \varepsilon_{33}^X E_{1s} \tag{10}$$

For convenience, we separated constitutive equations for electrically excited center part and side load part. In above equations, the subscripts "$c$" and "$s$" denotes center part and side part, respectively. Note that these equations are written with respect to the effective coordinate shown in Figure 2. Now, consider the following harmonic vibration equation, which is given by:

$$\rho \frac{\partial^2 u(x,t)}{\partial t^2} = \frac{\partial X_1}{\partial x} \tag{11}$$

$$u(x,t) = u(x) e^{j\omega t} \tag{12}$$

For the center part, by using constitutive equations and harmonic vibration equations, it is obtained:

$$\frac{\partial^2 u_{1c}}{\partial t^2} = \frac{1}{\rho s_{11}^E} \frac{\partial^2 u_{1c}}{\partial x^2} \tag{13}$$

This expression is the same as "wave differential equation" that can be found elsewhere. The velocity of the wave for the above equation is define as:

$$v_{11}^E = \sqrt{1/\rho s_{11}^E} \tag{14}$$

where $v_{11}^E$ is sound velocity related to compliance $s_{11}^E$. The general solution of the Eq. (14) for $u(x)$ would be in terms of sine and cosine function, which is of the form:

$$u_{1c}(x) = A_c \sin\left(\frac{\omega}{v_{11}^E} x\right) + B_c \cos\left(\frac{\omega}{v_{11}^E} x\right) \tag{15}$$

Since the displace function is asymmetric, but 0 at the center (center of gravity), cosine term should vanish. Therefore, for center part, we have the following equation of displacement:

$$u_{1c}(x) = A_c \sin\left(\frac{\omega}{v_{11}^E} x\right) \tag{16}$$

Where $A_c$ is constant. The wave equation for the side part should be considered separately. Due to different electrical boundary conditions, OC and SC, and SE should be separately considered. For the side part of OC and SC, we have $D$- constant conditions. In the similar manner, it is obtained:

$$\frac{\partial^2 u_{1s}}{\partial t^2} = \frac{1}{\rho s_{33}^D} \frac{\partial^2 u_{1s}}{\partial x^2} \tag{17}$$

The solution to the above wave equation is given by:

$$u_{1s}(x) = A_s \sin\left(\frac{\omega}{v_{33}^D} x\right) + B_s \cos\left(\frac{\omega}{v_{33}^D} x\right) \tag{18}$$

where $v_{33}^D$ is sound velocity related to compliance $s_{33}^D$, and $A_s$ and $B_s$ are constants. The side part does not have center of gravity, so it does not vanish at $x = 0$. For SE configuration, $E$- constant condition should be considered. In similar manner, it is obtained:

$$u_{1s}(x) = A_s \sin\left(\frac{\omega}{v_{33}^E} x\right) + B_s \cos\left(\frac{\omega}{v_{33}^E} x\right) \tag{19}$$

**B. Electrical Boundary Conditions**

For $k_{33}$ PE OC and SC, $D$- constant condition should be considered. For OC, assuming voltage applied to the center part is small, it can be assumed that $D_{1s}$ is zero. For SC, the expression for $D_{1s}$ should be derived according to the voltage constraint. Because of the short circuit condition, the potential difference between $x = al/2$ and $x = l/2$ should be zero even though the electric field distributes in between:

$$V_{1s} = -\int_{\frac{al}{2}}^{\frac{l}{2}} E_{1s}\, dx = 0 \tag{20}$$

By using constitutive equations and Eq. (20), and assuming that $D_{1s}$ is constant along the length direction, the following relationship can be obtained:

$$\int_{\frac{al}{2}}^{\frac{l}{2}} D_{1s}\, dx = D_{1s}\left(\frac{l}{2} - \frac{al}{2}\right) \tag{21}$$

With Eq. (21), the expression for $D_{1s}$ can be obtained from Eq. (10) and (18), taking into account the relationship between displacement and stress, which is shown in Eq. (11):

$$D_{1s} = \frac{2d_{33}\left[A_s\left\{\sin\left(\frac{l\omega}{2v_{33}^D}\right) - \sin\left(\frac{al\omega}{2v_{33}^D}\right)\right\} + B_s\left\{\cos\left(\frac{l\omega}{2v_{33}^D}\right) - \cos\left(\frac{al\omega}{2v_{33}^D}\right)\right\}\right]}{s_{33}^E l(1-a)} \tag{22}$$

For simplicity, let us express $D_{1s}$ with the following equation:

$$D_{1s} = \frac{d_{33}}{s_{33}^E} x_{1s,ave} \tag{23}$$

Where, $x_{1s,ave}$ is average value of strain $x_{1s}$ over the range of $x = al/2$ and $x = l/2$.

For $k_{33}$ SE, $E$- constant condition should be considered. Furthermore, due to charge neutralization originated from surface free charges that compensate surface bound charges, $E_{1s}$ should be zero.

## C. Mechanical Boundary Conditions

The mechanical boundary conditions are universal for all the PE configurations. First, the displacement should be continuous at $x = al/2$. By Equating Eq. (16) to Eq. (18), the following equations are obtained for the displacement continuation for OC and SC:

$$A_c \sin\left(\frac{\omega}{v_{11}^E}\frac{al}{2}\right) = A_s \sin\left(\frac{\omega}{v_{33}^D}\frac{al}{2}\right) + B_s \cos\left(\frac{\omega}{v_{33}^D}\frac{al}{2}\right) \quad (24)$$

For SE, the equation can simply be obtained by equating Eq. (16) and (19):

$$A_c \sin\left(\frac{\omega}{v_{11}^E}\frac{al}{2}\right) = A_s \sin\left(\frac{\omega}{v_{33}^E}\frac{al}{2}\right) + B_s \cos\left(\frac{\omega}{v_{33}^E}\frac{al}{2}\right) \quad (25)$$

Second, the stress should be continuous at $x = al/2$. The stress of the center part at $x = al/2$ for all PE configurations is obtained by using Eq. (11) and (16):

$$X_{1c} = \frac{\omega}{s_{11}^E v_{11}^E} A_c \cos\left(\frac{al\omega}{2v_{11}^E}\right) - \frac{d_{31}}{s_{11}^E} E_{3c} \quad (26)$$

The stress of the side part of $k_{33}$ PE OC, SC and SE can be similarly derived by utilizing mechanical wave equations. For $k_{33}$ OC, since $D_{1s} = 0$ was assumed, we obtain:

$$X_{1s} = A_s \frac{\omega}{v_{33}^D s_{33}^D} \cos\left(\frac{al\omega}{2v_{33}^D}\right) - B_s \frac{\omega}{v_{33}^D s_{33}^D} \sin\left(\frac{al\omega}{2v_{33}^D}\right) \quad (27)$$

For $k_{33}$ PE SC, by using Eq. (23), we obtain:

$$X_{1s} = A_s \frac{\omega}{v_{33}^D s_{33}^D} \cos\left(\frac{al\omega}{2v_{33}^D}\right) - B_s \frac{\omega}{v_{33}^D s_{33}^D} \sin\left(\frac{al\omega}{2v_{33}^D}\right) - \frac{k_{33}^2 x_{1s,ave}}{s_{33}^D} \quad (28)$$

For $k_{33}$ PE SE, by applying $E_{1s} = 0$, we obtain:

$$X_{1s} = A_s \frac{\omega}{v_{33}^E s_{33}^E} \cos\left(\frac{al\omega}{2v_{33}^E}\right) - B_s \frac{\omega}{v_{33}^E s_{33}^E} \sin\left(\frac{al\omega}{2v_{33}^E}\right) \quad (29)$$

Then stress continuation condition should be achieved by equating Eq. (26) to Eq. (27), (28) and (29).

Third, the stress should be zero at each edge of the bar, which is located at $x = l/2$. The equations can simply be expressed by using Eq. (27), (28) and (29), but using $x = l/2$ rather than $x = al/2$ For OC, we have:

$$A_s \frac{\omega}{v_{33}^D s_{33}^D} \cos\left(\frac{l\omega}{2v_{33}^D}\right) - B_s \frac{\omega}{v_{33}^D s_{33}^D} \sin\left(\frac{l\omega}{2v_{33}^D}\right) = 0 \quad (30)$$

For SC, we have:

$$A_s \frac{\omega}{v_{33}^D s_{33}^D} \cos\left(\frac{l\omega}{2v_{33}^D}\right) - B_s \frac{\omega}{v_{33}^D s_{33}^D} \sin\left(\frac{l\omega}{2v_{33}^D}\right) - \frac{k_{33}^2 x_{1s,ave}}{s_{33}^D} = 0 \quad (31)$$

For SE, we have:

$$A_s \frac{\omega}{v_{33}^E s_{33}^E} \cos\left(\frac{l\omega}{2v_{33}^E}\right) - B_s \frac{\omega}{v_{33}^E s_{33}^E} \sin\left(\frac{l\omega}{2v_{33}^E}\right) = 0 \quad (32)$$

By this moment, for each configuration, we have three equations from three mechanical boundary conditions and three unknown constants ($A_c$, $A_s$ and $B_s$). Solving the system of equations will provide the solution to these constants. Because only the center part is measured, only $A_c$ is required to solve the final admittance equations.

D. Solving Admittance Equation

For PE, since we measure the current (admittance) through the center electrode, we have:

$$I = j\omega \iint D_{3c} dx dy = 2j\omega w \left(\int_0^{\frac{al}{2}} D_{3c} dx\right) \quad (33)$$

Then using constitutive equations, it is obtained:

$$I = 2j\omega w \left[\frac{A_c d_{31}}{s_{11}^E} \sin\left(\frac{al\omega}{2v_{11}^E}\right) + \frac{al\varepsilon_0 \varepsilon_{33}^X \left(1 - k_{31}^2\right) E_{3c}}{2}\right] \quad (34)$$

The expression for admittance is simply given by:

$$Y = -\frac{I}{V} = \frac{I}{E_{3c}t} \tag{35}$$

Finally, the expression for the admittance for all the configuration is given by:

$$Y_{OC} = jw \left[ \frac{2d_{31}^{*2} v_{11}^{E*} v_{33}^{D*} s_{33}^{D*} \cos\left(\frac{\omega}{v_{33}^{D*}} \frac{(a-1)l}{2}\right) \sin\left(\frac{\omega}{v_{11}^{E*}} \frac{al}{2}\right)}{t s_{11}^{E*} v_{33}^{E*} s_{33}^{E*} \cos\left(\frac{\omega}{v_{33}^{D*}} \frac{(a-1)l}{2}\right) \cos\left(\frac{\omega}{v_{11}^{E*}} \frac{al}{2}\right) + v_{11}^{E*} s_{11}^{E*} \sin\left(\frac{\omega}{v_{11}^{E*}} \frac{al}{2}\right) \sin\left(\frac{\omega}{v_{33}^{D*}} \frac{(a-1)l}{2}\right)} + \frac{al\omega \varepsilon_0 \varepsilon_{33}^{X*}(1-k_{31}^{*2})}{t} \right] \tag{36}$$

$$Y_{SC} = jw \left[ \frac{2d_{31}^{*2} v_{11}^{E*} v_{33}^{D*} s_{33}^{D*}}{s_{11}^{E*} t \left( \frac{s_{33}^{D*} v_{33}^{D*}}{\tan\left(\frac{al\omega}{2v_{11}^{E*}}\right)} + v_{11}^{E*} s_{11}^{E*} \left\{ \frac{8d_{33}^{*2} v_{33}^{D*} \sin^2\left(\frac{(1-a)l\omega}{4v_{33}^{D*}}\right) + l\omega(1-a)\varepsilon_0 \varepsilon_{33}^{X*} s_{33}^{E*} \sin\left(\frac{(1-a)l\omega}{2v_{33}^{D*}}\right)}{2d_{33}^{*2} v_{33}^{D*} \sin\left(\frac{(1-a)l\omega}{2v_{33}^{D*}}\right) - l\omega(1-a)\varepsilon_0 \varepsilon_{33}^{X*} s_{33}^{E*} \cos\left(\frac{(1-a)l\omega}{2v_{33}^{D*}}\right)} \right\} \right)} + \frac{al\omega \varepsilon_0 \varepsilon_{33}^{X*}(1-k_{31}^{*2})}{t} \right] \tag{37}$$

$$Y_{SE} = jw \left[ \frac{2d_{31}^{*2} v_{11}^{E*} v_{33}^{E*} s_{33}^{E*} \cos\left(\frac{\omega}{v_{33}^{E*}} \frac{(a-1)l}{2}\right) \sin\left(\frac{\omega}{v_{11}^{E*}} \frac{al}{2}\right)}{t s_{11}^{E*} v_{33}^{E*} s_{33}^{E*} \cos\left(\frac{\omega}{v_{33}^{E*}} \frac{(a-1)l}{2}\right) \cos\left(\frac{\omega}{v_{11}^{E*}} \frac{al}{2}\right) + v_{11}^{E*} s_{11}^{E*} \sin\left(\frac{\omega}{v_{11}^{E*}} \frac{al}{2}\right) \sin\left(\frac{\omega}{v_{33}^{E*}} \frac{(a-1)l}{2}\right)} + \frac{al\omega \varepsilon_0 \varepsilon_{33}^{X*}(1-k_{31}^{*2})}{t} \right] \tag{38}$$

For the final admittance equations, it is noteworthy to see the superscript star (*) on the physical parameters, meaning the parameters are complex, including real and imaginary (loss) parts. These parameters can be written as:

$$\varepsilon_{33}^{X*} = \varepsilon_{33}^{X}(1 - j\tan\delta_{33}') \tag{39}$$

$$s_{11}^{E*} = s_{11}^{E}(1 - j\tan\phi_{11}') \tag{40}$$

$$s_{33}^{D*} = s_{33}^{D}(1 - j\tan\phi_{33}''') \tag{41}$$

$$s_{33}^{E*} = s_{33}^{E}(1 - j\tan\phi_{33}') \tag{42}$$

$$d_{31}^{*} = d_{31}(1 - j\tan\theta_{31}') \tag{43}$$

$$d_{33}^{*} = d_{33}(1 - j\tan\theta_{33}') \tag{44}$$

$$k_{31}^{*} = k_{31}(1 - j\tan\chi_{31}) \tag{45}$$

$$k_{33}^{*} = k_{33}(1 - j\tan\chi_{33}) \tag{46}$$

In Eq. (41), The corresponding loss of $s_{33}^{D}$ is not exactly the extensive elastic loss ($\tan\phi_{33}$), but it is described as triple-prime loss ($\tan\phi_{33}'''$) due to mechanical constraint difference with the $k_t$ mode. It is given by [10]:

$$\tan \phi'''_{33} = \frac{1}{Q_{B,33}} = \frac{1}{1-k_{33}^2}[\tan \phi'_{33} - k_{33}^2(2\tan \theta'_{33} - \tan \delta'_{33})] \quad (47)$$

$\tan \phi_{33}$ can rather be directly determined from thickness mode ($k_t$) specimen.

### E. Mathematical Limit of Analytical Solutions

So far, analytical solutions for $k_{33}$ PE are derived. One of validation methods for analytical solutions is to take limit of the admittance functions. If we take the center portion "$a$" as 100 % ($a \to 1$) for all the admittance equations, meaning the center part takes up the entire PE geometry, we obtain:

$$Y = j\omega \frac{\varepsilon_{33}^X wl}{t}\left[(1-k_{31}^2) + k_{31}^2 \frac{\tan(\omega l/2v_{11}^E)}{\omega l/2v_{11}^E}\right] \quad (48)$$

As a result of $a \to 1$, all the equations beautifully converge to that of Standard $k_{31}$ mode [7]. It is obvious from the fact that the center part is $k_{31}$ mode for all the PE geometry. Next, when "$a$" goes to zero ($a \to 0$), antiresonance frequency ($f_B$) of PE OC should approach to that of standard specimen, while resonance frequency ($f_A$) of PE SC should approach to that of standard sample, since PE OC corresponds to antiresonance behavior and PE SC corresponds to resonance behavior for $k_{33}$ mode. Figure 3 shows analytical admittance curves for standard sample [15] and for PE OC and SC with varying "$a$". In Figure 3, it clearly shows that the derived admittance curves for OC and SC approaches to $f_B$ and $f_A$, respectively, of standard $k_{33}$ mode. It should be note that PE OC case is more sensitive to the portion of "$a$" than PE SC case. Therefore, different from parameter determination method for $k_{31}$ PE, in which the parameters are calculated through the certain equations[5], $k_{33}$ PE analytical solutions are required to fit into corresponding experimental data in order to determine $k_{33}$ mode-related parameters.

### IV. Comparison with Finite Element Analysis (FEA)

The mathematical limit partially validates analytical solutions; they should be compared to both finite element analysis (FEA) for definite evaluation. Figure 4 shows direct comparisons of admittance curves of PE OC, SC and SE between analytical solution and FEA. For FEA simulation, the parameters of PZT 5A, which are already built in the software, were utilized. The physical parameters and losses of PZT 5A are listed in Table 1. It should be noted that due to loss isotropy in ATILA FEA software, the subscripted indices are denoted as (i,j). For the calculation, the piezoelectric loss was set to be 1.6 %, about the average of the sum of dielectric and elastic loss. The discrepancies between admittance curves from analytical solution and ATILA FEA is mainly due to two reasons: (1) 3D analysis in ATILA, whereas 1D analysis in the analytical solutions. Due to 3D consideration of FEA, the peak heights of admittance curves from FEA are smaller than those of admittance curves from analytical solutions. (2) FEA's consideration of higher harmonic generation, while analytical solution only considers fundamental resonance. Due to this difference, the base lines far from resonance and antiresonance frequencies of admittance curves deviate. Despite these discrepancies, only about 0.1 % deviation occurs in $f_A$ and $f_B$, and about 2 % deviation occurs in $Q_A$ and $Q_B$. These values are especially important, since they are directly related to physical parameters and loss determination.

## V. Fitting of analytical solutions to experimental data

After analytical solutions were validated with the aid of FEA, they were fitted to experimental data. Since the center part is $k_{31}$ mode, the related physical parameters were measured with standard $k_{31}$ sample. Determined parameters of $k_{31}$ mode are listed in Table 2. The parameter determination method for $k_{31}$ mode follows the one in Zhuang *et al*. [8]. For curve fitting, in order to simplify and minimize the fitting parameters, the following method was utilized: first, $\rho$ was determined by measuring the mass of each sample. Also, "$a$", the center portion, was measured with the help of optical microscope. Although we designed $a$ as 10 % of the total length, about ± 20 % deviation occurred. Therefore, it is recommended to directly measure $a$ for fitting with less error. By using the center portion, $\varepsilon_{33}^X$ and $\tan\delta'_{33}$ were measured in off-resonance regime (at 1 kHz), from capacitance and phase lag, respectively. Then, by using OC specimen, $s_{33}^D$ and $\tan\phi'''_{33}$ were determined with fitting. It should be noted that the fringing "output" electric field occurring in the side of OC can overestimate the value of $s_{33}^D$, as it occurs in standard specimen. Therefore, by utilizing the finite element results in our paper [11], $s_{33}^D$ value was calibrated by considering aspect ratio of the sample. Next, by using SE specimen, $s_{33}^E$ and $\tan\phi'_{33}$ were determined with fitting. Finally, using already determined parameters from the previous process, $d_{33}$ and $\tan\theta'_{33}$ were determined by using SC sample. It should be noted that $d_{31}$, which was determined in advance by measuring standard $k_{31}$ sample, was underestimated by about – 10 % to generate good fittings. The reason may be attributed to the imperfect poling issue at the boundary between the center and side portions. However, varying $d_{31}$ in analytical solutions does not have significant effect on $f_A$, $f_B$, $Q_A$ and $Q_B$; rather, $d_{31}$ is just the "actuator" parameter, not the key "side-load" parameter. Since all the parameters required to be determined from fitting ($s_{33}^D$, $\tan\phi'''_{33}$, $s_{33}^E$, $\tan\phi'_{33}$, $d_{33}$ and $\tan\theta'_{33}$) depend on $f_A$, $f_B$, $Q_A$ and $Q_B$, the values of determined parameters can be considered valid. Figure 5 shows experimental data fitted to analytical solutions. Considering the underestimation of $d_{31}$ during the fitting process, analytical solution gives out excellent fitting results, with the percentage error of less than 1 %.

## VI. Discussion

After all possible fitting parameters are determined, the real part of Eq. (46), which is called electromechanical coupling factors ($k_{33}$) can be calculated in two ways. With $\varepsilon_{33}^X$ determined from off-resonance measurement, $d_{33}$ determined from SC specimen and $s_{33}^E$ determined from SE specimen, it can be obtained:

$$k_{33}^2 = \frac{d_{33}^2}{s_{33}^E \varepsilon_0 \varepsilon_{33}^X} \tag{49}$$

Also, from $s_{33}^D$ determined from OC specimen and $s_{33}^E$ determined from SE specimen, it can be calculated:

$$(1 - k_{33}^2) = \frac{s_{33}^D}{s_{33}^E} \tag{50}$$

The imaginary part of electromechanical coupling factor ($\tan \chi_{33}$) are defined as "coupling loss", which is given by:

$$\tan \chi_{33} = 2 \tan \theta_{33}' - \tan \delta_{33}' - \tan \phi_{33}' \tag{51}$$

The determined values of physical parameters and losses from curve fitting are shown in Table 3. The fitting method can be validated by comparing $k_{33}$ value determined from two different methods, by following Eq. (49) and (50). $k_{33}$ determined from Eq. (49) is 70.3 % and that determined from Eq. (50) is 69.6 %; these are in good agreement.

## VII. Conclusion

In summary, the "Partial Electrode" configuration for $k_{33}$ mode was introduced in order to resolve issues in the IEEE Standard $k_{33}$ mode specimen, which can determine experimentally both "intrinsic" and "extrinsic" physical parameters in piezo-ceramics. The analytical solutions for $k_{33}$ PE have been derived, with the aid of piezoelectric constitutive equations and electrical/mechanical boundary conditions. The derived admittance curves from analytical solutions well match with those simulated from ATILA FEA, in terms of resonance frequencies and mechanical quality factors. After analytical solutions are validated through the comparison, they were fitted to real experimental data to determine physical parameters of $k_{33}$ mode. The fitting method was confirmed by comparing $k_{33}$ values determined from two different equations. The fitting method to determine physical parameters and losses with analytical solutions will provide more accurate parameter determination compared to the method using the IEEE standard $k_{33}$ specimen by providing much larger capacitance with less fringing electric field effect, free of specimen setup issue, and direct determination of intensive elastic compliance and losses.

**Acknowledgement**

This work was supported by Office of Naval Research under Grant Number N00014-17-1-2088

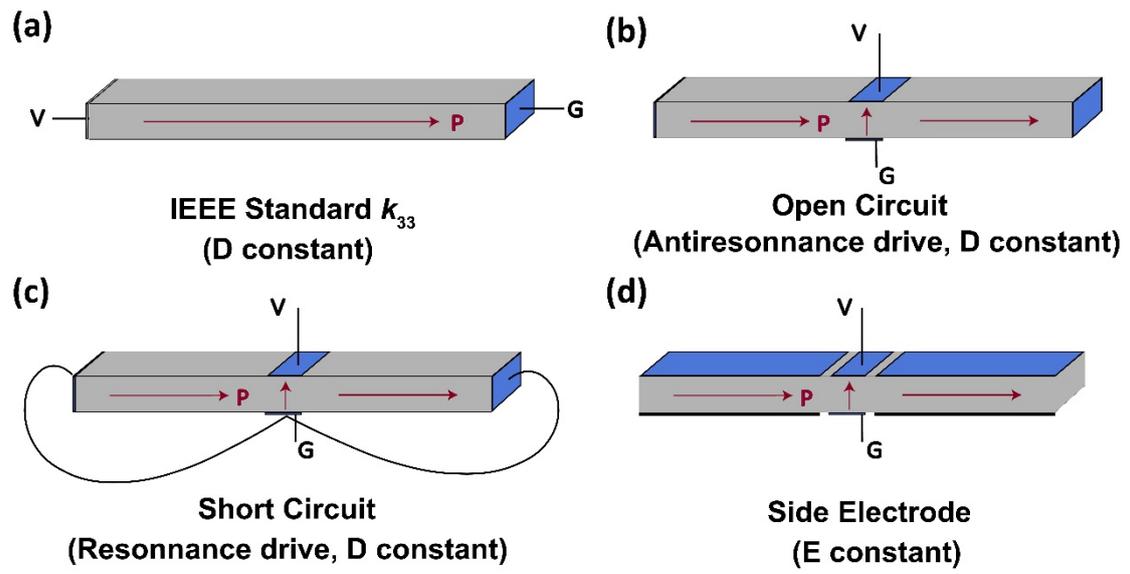

Figure 1. The schematic diagrams of (a) IEEE Standard $k_{33}$ mode piezoelectric specimen, (b) PE open circuit (OC), (c) PE short circuit (SC), and (d) PE side electrode (SE)

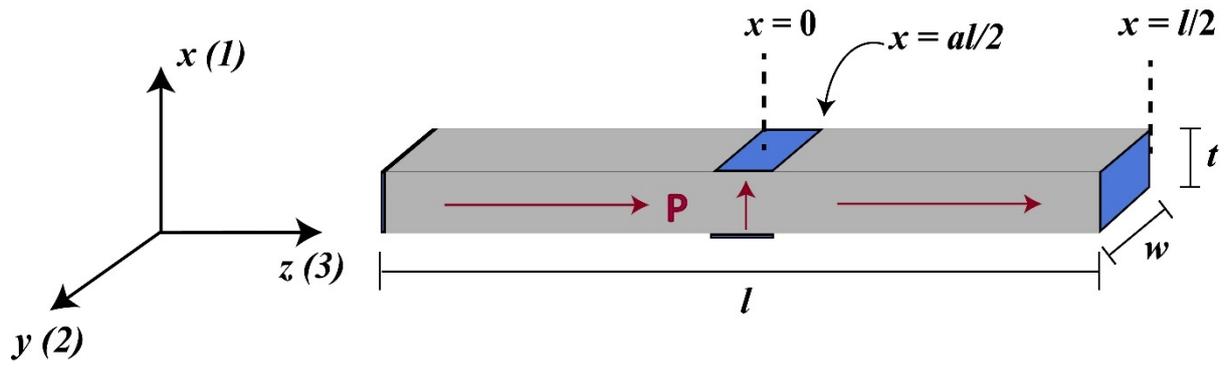

Figure 2. Defined orientation and dimension of PE used in derivation process of analytical solutions.

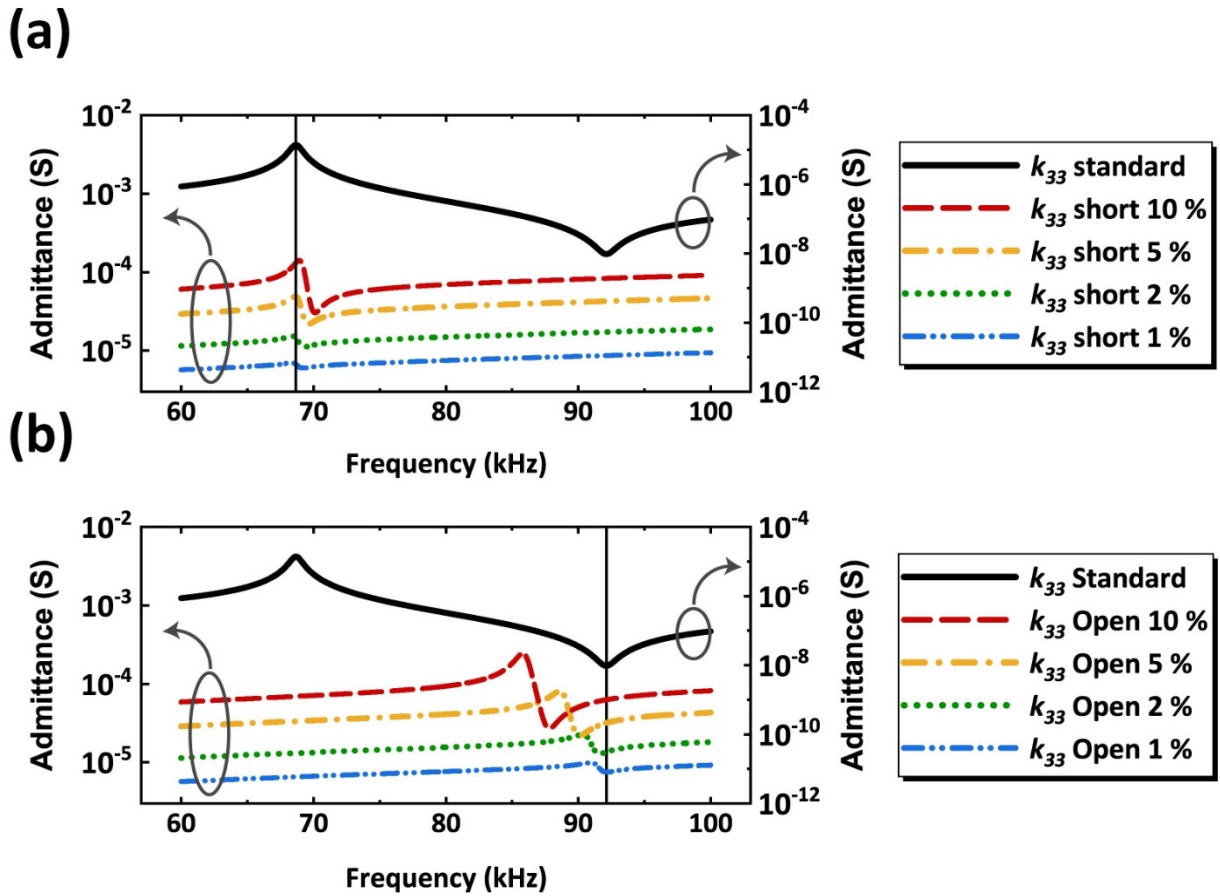

Figure 3. (a) Frequency dependence of analytical admittance curve plot for standard and SC specimen for resonance frequency comparison and (b) analytical admittance curve plot for standard and SC specimen for antiresonance frequency comparison with the center portion of 10 % (red dashed), 5 % (yellow dot-dashed), 2 % (green dotted), and 1 % (blue double dot-dashed). Black vertical lines in (a) and (b) shows resonance and antiresonance frequency of the admittance curve of standard specimen.

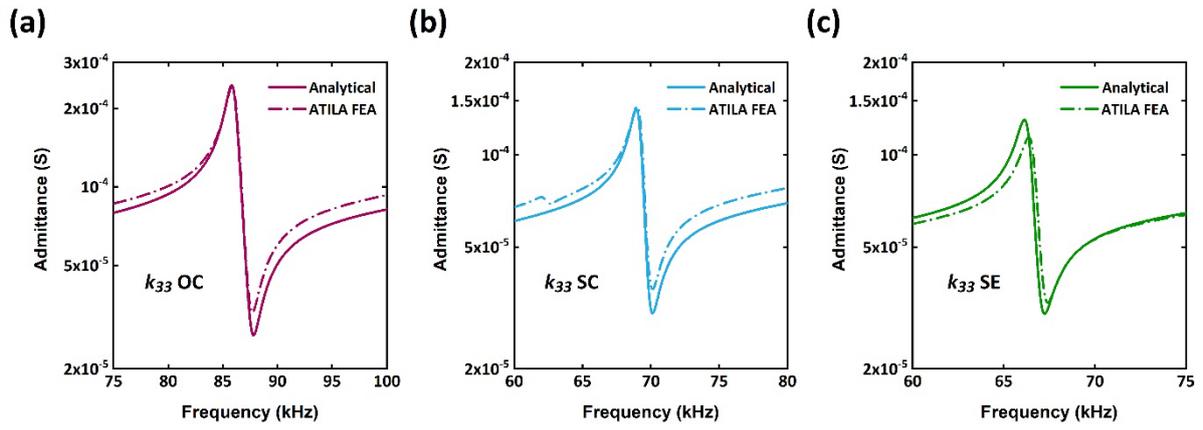

Figure 4. Comparison of admittance curves of analytical solutions and from ATILA FEA simulation for (a) OC, (b) SC, and (c) SC.

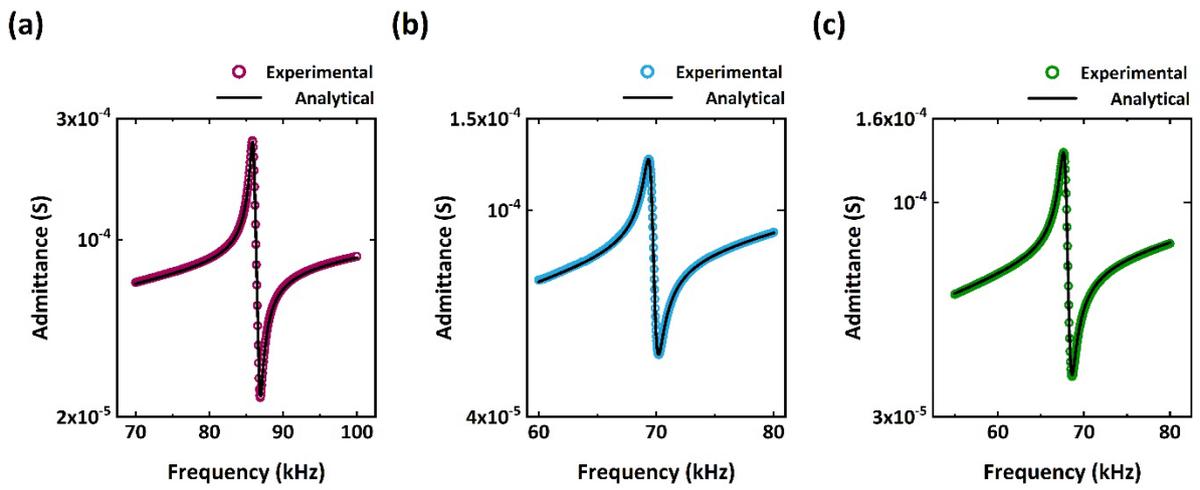

Figure 5. Experimental admittance curves of PIC 255 with (a) OC (Purple open circle), (b) SC (Sky blue open circle) and (c) SE configuration (Green open circle). Black solid line shows analytical admittance curves fitted to experimental data.

**Physical properties of PZT 5A used in ATILA FEA simulation**

| Real Parameters | | | | | |
|---|---|---|---|---|---|
| $\rho$ (kg/m³) | $\varepsilon_{33}^X$ …. | $s_{11}^E$ (μm²/N) | $s_{33}^E$ (μm²/N) | $d_{31}$ (pC/N) | $d_{33}$ (pC/N) |
| 7750 | 1700 | 16.4 | 18.8 | -171 | 374 |

| Imaginary Parameters (losses) | | |
|---|---|---|
| $\tan \delta'_{ij}$ (%) | $\tan \phi'_{ij}$ (%) | $\tan \theta'_{ij}$ (%) |
| 2 | 1.3 | 1.6 |

Table 1. Physical parameter inputs for PZT 5A that was used for ATILA FEA simulation. Due to isotropic consideration of losses in ATILA, the subscripted indices in losses are represented as (i,j).

## Parameters Determined from Standard $k_{31}$ Specimen (PIC 255)

| Real Parameters | | | |
|---|---|---|---|
| $\varepsilon_{33}^X$ | $s_{11}^E$ | $d_{31}$ | $k_{31}$ |
| …. | (μm²/N) | (pC/N) | (%) |
| 1808 ± 0.1 % | 16.1 ± 0.1 % | -187 ± 4 % | 0.37 ± 4 % |

| Imaginary Parameters (losses) | | | |
|---|---|---|---|
| $\tan \delta_{33}'$ | $\tan \phi_{11}'$ | $\tan \theta_{31}'$ | $\tan \chi_{31}$ |
| (%) | (%) | (%) | (%) |
| 1.6 ± 1 % | 1.2 ± 2 % | 2.2 ± 6 % | 1.6 ± 16 % |

Table 2. Physical parameters determined from standard $k_{31}$ specimen

**Parameters Determined from PE Sample**

| Real Parameters | | | | | |
|---|---|---|---|---|---|
| $\varepsilon_{33}^X$ .... | $s_{33}^D$ (μm²/N) | $s_{33}^E$ (μm²/N) | $d_{33}$ (pC/N) | $k_{33}$ from Eq. (49) (%) | $k_{33}$ from Eq. (50) (%) |
| 1834 | 9.18 | 17.8 | 378 | 70.3 | 69.6 |

| Imaginary Parameters (losses) | | | | |
|---|---|---|---|---|
| $\tan \delta_{33}'$ (%) | $\tan \phi_{33}'''$ (%) | $\tan \phi_{33}'$ (%) | $\tan \theta_{33}'$ (%) | $\tan \chi_{33}$ (%) |
| 1.6 | 0.56 | 1.2 | 1.8 | 0.8 |

Table 3. Parameters determined from curve fitting of PE samples with derived analytical solutions.